\documentclass
[notitlepage,longbibliography,superscriptaddress,pre,nofootinbib]{revtex4-2}%
\usepackage{url}
\usepackage{placeins}
\usepackage{tikz}
\usepackage{booktabs}
\usepackage{pbox}
\usepackage{color}
\usepackage{relsize}
\usepackage{graphicx}
\usepackage[intlimits]{amsmath}
\usepackage{amsxtra, amssymb,fancyhdr, amsthm,latexsym}
\usepackage{cases}
\usepackage{verbatim}
\usepackage{float}
\usepackage[hidelinks]{hyperref}
\usepackage{textcomp}
\usepackage{amsfonts}
\usepackage{amssymb}
\usepackage[mathlines]{lineno}%
\providecommand{\U}[1]{\protect\rule{.1in}{.1in}}
\usetikzlibrary{automata,arrows,positioning,calc}

\begin{document}
\begin{empty}
\title{Bacterial aggregation in the presence of directed motion}
\author{Timothy Power}
\email{Timothy.Power@dal.ca}
\affiliation{Dalhousie University Halifax, Nova Scotia, Canada}
\author{Hildur Knutsdottir}
\email{hildurknuts@gmail.com}
\affiliation{University of British Columbia, Vancouver, Canada}
\author{Leah Edelstein-Keshet}
\email{keshet@math.ubc.ca}
\affiliation{University of British Columbia, Vancouver, Canada}
\author{Theodore Kolokolnikov}
\email{tkolokol@gmail.com}
\affiliation{Dalhousie University Halifax, Canada}
\begin{abstract}
We model directed motion of cells that aggregate into a cluster in the
presence of chemotaxis or fluid transport. Starting from a stochastic
agent-based model (ABM) of Galante \& Levy (2013), Chavy-Waddy \& Kolokolnikov (2016), we consider two
model variants, with (I) directed motion of individuals, or (II) implicit
individual bias in one direction. In both cases, we derive fourth order
nonlinear PDEs representing the continuum limits of the models as a
generalization of  Chavy-Waddy \& Kolokolnikov (2016). An exact analytic solution then
describes cluster density profile as well as the group velocity, and we identify
the regimes where the travelling cluster exists. Excellent
agreement is observed between the discrete models and the exact
solutions of the PDEs in both models, including the existence and speed of the
travelling cluster.
\end{abstract}
\maketitle
\end{empty}

\section{Introduction}

In a series of papers \cite{msel1, msel2, galante2011stochastic, smsp,
chavy2016local}, the authors proposed several models of motion of
cyanobacteria \emph{Synechocystis sp.} Two key aspects were
emphasized:\ aggregation -- localized areas of high bacterial density; and the
motion towards a light source (phototaxis). While phototaxis is induced by a
light gradient, bacterial aggregation can happen even in the absence of such
gradient \cite{msel1,chavy2016local}.

Models for collective motion of cells date back to the 1970's with the classic
Keller-Segel model \cite{keller1971traveling} for traveling bands of
chemotactic bacteria, moving up a nutrient gradient. In
\cite{lauffenburger1984traveling}, the approximate shape of the cell density
profile was derived from a similar continuum model. More recently, chemotactic
E. coli were tracked experimentally in micro channels and compared with a
one-dimensional (1D) continuum model \cite{saragosti2011directional}. Indeed,
the development of micro fluidic devices \cite{kim2012microfluidics} and
adhesive patterns has enhanced the ability to measure cell motion and cell
interaction in simplified system geometry. Such experiments also extend to
mammalian cells, though typically with smaller cell populations. Chemotaxis
and cell pair interactions were quantified in 1D microfluidic device by
\cite{lin2015interplay}, and interactions of ``trains" of cells on 1D adhesive
strips were observed in \cite{jain2020role}. Observations in vivo of embryo
development reveal that neural crest cells become polarized, attract one
another, and migrate to target sites with directional preference
\cite{leonard2020road,mclennan2015neural,schumacher2016multidisciplinary}. In
some cases, e.g. the development of the lateral line in zebrafish, a group of
roughly 100 cells migratess along a 1D track in a polarized cluster
\cite{mehes2014collective}. Such studies motivate our goal of understanding
how the motion of cells is influenced by external bias, cell-cell
interactions, fluid flow in the environment, and directional cues. In contrast
to other modeling studies where multiple cells types and signals are
considered, we here consider a minimal 1-cell-type model with basic cell-cell
interactions to determine conditions for cohesive group motion.

Numerous other models of bacterial aggregation have been proposed. There is a
very large literature on this topic, often referred to as collective motion;
here we mention just a few representative studies. This includes Langevin-type
nonlocal models, \cite{falco2023local, buttenschon2024cells, mogilner1999non,
wehlitz2025approximating}, reaction-diffusion models
\cite{kawasaki1997modeling, wang2007classical}, cellular automata models
\cite{ermentrout1993cellular, deutsch2000new}, state-dependent diffusion
\cite{liu2011sequential}, and correlated random walk models
\cite{eftimie2012hyperbolic, buttenschon2019correlated}.

In \cite{chavy2016local} the authors derived a fourth order nonlinear PDE in
the continuum limit, which exhibits localized spikes corresponding to the
clusters observed in the discrete model from \cite{msel1}, in the absence of
light gradient. In this paper we extend the work from \cite{chavy2016local} by
introducing two different mechanisms that induce the cluster motion. These are:

\begin{itemize}
\item[(I)] A transport velocity, for example, due to fluid flow.

\item[(II)] Biased direction selection, such as a bias towards a source of
light
or a chemical gradient.
\end{itemize}

For both models, we analytically derive the aggregation profile using a
variant of the method first proposed in \cite{chavy2016local}. In addition to
the profile shape, this method also yields an exact analytical expression for
the speed of spike motion, as a function of model parameters. We further
determine the parameter regime in which such a travelling cluster exists
including a region of bistability in model II\ where a homogeneous and cluster
solution coexist.

\section{Model I:\ directed motion\label{sec:1}}

At an agent-based level, the model description is as follows. Each individual
has three components of motion

\begin{enumerate}
\item Random: with a rate of $2a\cdot dt,$ a cell
moves one step either left or right.

\item Aggregation: With a rate of $b\cdot dt,$ the
cell randomly chooses a neighbour within its sensing radius $d$.
It then moves one step towards that neighbour.

\item Directed motion:\ with a rate $v\cdot dt$, the cell
is transported one step to the right.
\end{enumerate}

Note that this model (with $v=0$) is a simplification of the model introduced
in \cite{msel1, chavy2016local}. There, each particle remembers its past
direction. By contrast here, we consider a simpler model where a new direction
is chosen at every time-step.

\begin{figure}[ptb]
\begin{center}
\includegraphics[width=1\textwidth]{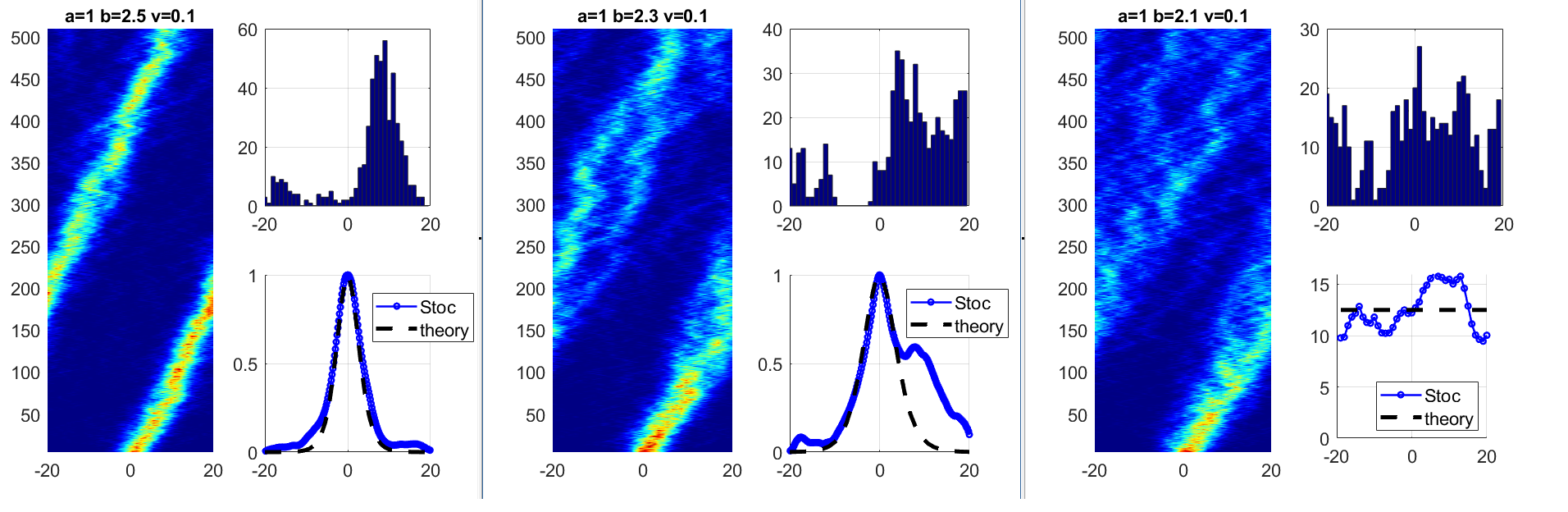}
\end{center}
\caption{Three distinct simulations of the stochastic ABM version of Model I
on a periodic domain with $n=1000$ particles. and for parameters as given. For
each of three panels, left is a\ time-space plot (kymograph) of the particle
density. Top right is a histogram of particle locations at $t=500.$ Bottom
right shows the average profile of the moving cluster (histogram averaged over
$t=300\ldots500$), normalized to center the distribution at $x=0$. Overlaid is
the theoretical prediction given in (\ref{uas}). In all simulations we took
$d=1.$ For the third simulation, the aggregation parameter $b$ is exactly at
the threshold $b_{c}=2a+v$ and the cluster dissipates.}%
\label{fig:stoch}%
\end{figure}

It was previously observed that for sufficiently large aggregation rate $b,$
the particles start to cluster \cite{msel1, chavy2016local}. Conversely, we
find that the cluster starts to dissipate if either the velocity $v$ or random
motion rate $2a$ is increased too much. A sample simulation demonstrating
these effects within the stochastic model above is shown in Figure
\ref{fig:stoch}. 

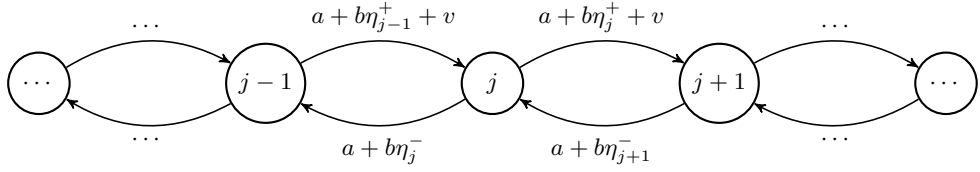
\begin{figure}[ptb]
\begin{center}
\begin{tikzpicture}[->, >=stealth', auto, semithick, node distance=3cm]
\tikzstyle{every state}=[fill=white,draw=black,thick,text=black,scale=1]
\node[state]    (j)                     {$j$};
\node[state]    (jp)[right of=j]         {$j+1$};
\node[state]    (jpp)[right of=jp]        {$\cdots$};
\node[state]    (jm)[left of=j]         {$j-1$};
\node[state]    (jmm)[left of=jm]        {$\cdots$};
\path
(jm) edge[bend left]     node{$a+b\eta^+_{j-1}+v$}         (j)
(jm) edge[bend left]     node{$\cdots$}         (jmm)
(jmm) edge[bend left]     node{$\cdots$}         (jm)
(jp) edge[bend left]     node{$\cdots$}         (jpp)
(jpp) edge[bend left]     node{$\cdots$}         (jp)
(jp) edge[bend left]     node{$a+b\eta^-_{j+1}$}         (j)
(j)  edge[bend left]     node{$a+b\eta^+_j+v$}         (jp)
(j)  edge[bend left]     node{$a+b\eta^-_j$}         (jm);
\end{tikzpicture}
\end{center}
\caption{Transitions between states that lead to the ODE system (\ref{ode}).
The fractions $\eta_{j}^{\pm}$ are defined in (\ref{eta}).}%
\label{fig:transition}%
\end{figure}

In the limit of a large number of particles, we discretize the space into $N$
bins and let $U_{j}$ be the particle density within the bin $j.$ Figure
\ref{fig:transition} shows the transition diagram between different bins.
Taking the limit $dt\rightarrow0$, the agent-based model reduces to an ODE
system,
\begin{align}
\frac{dU_{j}}{dt}  &  =a\left(  U_{j-1}-2U_{j}+U_{j+1}\right)  +b\left(
\eta_{j+1}^{-}U_{j+1}-U_{j}+\eta_{j-1}^{+}U_{j-1}\right)  +v\left(
U_{j-1}-U_{j}\right)  ;\label{ode}\\
\eta_{j}^{\pm}  &  =\frac{\sum_{k=1}^{d}U_{j\pm k}}{\sum_{k=1}^{d}%
(U_{j+k}+U_{j-k})}. \label{eta}%
\end{align}
Here, $\eta_{j}^{-}$ and $\eta_{j}^{+}$ correspond to the fraction of
particles either to the left (-) or to the right (+) of bin $j,$ within a
sensing distance $d.$

In the limit of a large number of bins, the model (\ref{ode}) reduces to a PDE
that we now derive. Approximate $U_{j}(t)\sim u\left(  x,t\right)  $ where
$x=jh$, and $h$ is the spacing between the bins, assumed to be small. Using a
Taylor series expansion in $h$, following the procedure in
\cite{chavy2016local}, expanding to two orders we obtain the approximation%

\begin{equation}
u_{t}=-\frac{h^{2}}{2}(bd-2a-v)u_{xx}(x,t)-vhu_{x}(x,t). \label{diffusion}%
\end{equation}
Note that the coefficient of $u_{xx}$ changes sign at the critical velocity%
\begin{equation}
v_{c}:=bd-2a. \label{vc}%
\end{equation}
The uniform steady state is stable provided that $v>v_{c}$. In the opposite
case when $0\leq v<v_{c}$ (or equivalently, $bd>2a+v$), the diffusion
coefficient is negative and so diffusion equation (\ref{diffusion}%
)\ ill-posed (unstable). In this case, higher order terms (up to fourth order)
in the expansion are needed to capture the aggregation behaviour. After
performing this expansion following the procedure in \cite{chavy2016local}, we
obtain a PDE approximation%

\begin{equation}
u_{t}=-Au_{xx}(x,t)-Bu_{xxxx}+C\left(  \frac{u_{x}u_{xx}}{u}\right)
_{x}-vhu_{x}(x,t)-v\frac{h^{3}}{6}u_{xxx}, \label{pde0}%
\end{equation}
where
\begin{equation}
A=\frac{h^{2}}{2}(bd-2a-v),\ B=\frac{h^{4}}{24}\left(  b(d^{3}+2d^{2}%
+3d+1)-2a-v\right)  ,\quad\text{and}\ C=\frac{h^{4}}{24}b(2d+1)(d+1)^{2}.
\end{equation}
Equation (\ref{pde0})\ reduces to the model from \cite{chavy2016local} when
$v=0.$ Note that $B>0$ automatically whenever $A>0$, since $b(d^{3}%
+2d^{2}+3d+1)-2a-v>bd-2a-v>0.$ This fourth-order term effectively stabilizes
the high-frequency modes and makes the equation well-posed.

Following \cite{chavy2016local}, we non-dimensionalize (\ref{pde0})\ to make
the coefficients of $u_{xx}$ and $u_{xxxx}$ be unity. Namely, we scale space
and time as%
\begin{equation}
x=\sqrt{B/A}\tilde{x},\ \ \ \ t=\frac{B}{A^{2}}\tilde{t}. \label{xchange}%
\end{equation}
Dropping the tilde, we obtain the dimensionless PDE%
\begin{equation}
u_{\tilde{t}}=-u_{xx}-u_{xxxx}+\alpha\left(  \frac{u_{x}u_{xx}}{u}\right)
_{x}-\beta u_{x}-\gamma u_{xxx}, \label{pde}%
\end{equation}
where%
\begin{align}
\alpha &  =\frac{b(2d+1)(d+1)^{2}}{\left(  b(d^{3}+2d^{2}+3d+1)-2a-v\right)
},\nonumber\\
\beta &  =3^{-1/2}v\left(  \frac{b(d^{3}+2d^{2}+3d+1)-2a-v}{(bd-2a-v)^{3}%
}\right)  ^{1/2},\\
\gamma &  =\frac{2}{\sqrt{3}}v\frac{1}{\sqrt{\left(  b(d^{3}+2d^{2}%
+3d+1)-2a-v\right)  (bd-2a-v)}}.\nonumber
\end{align}
\bigskip

\begin{figure}[ptbh]
\begin{center}
\includegraphics[width=0.8\textwidth]{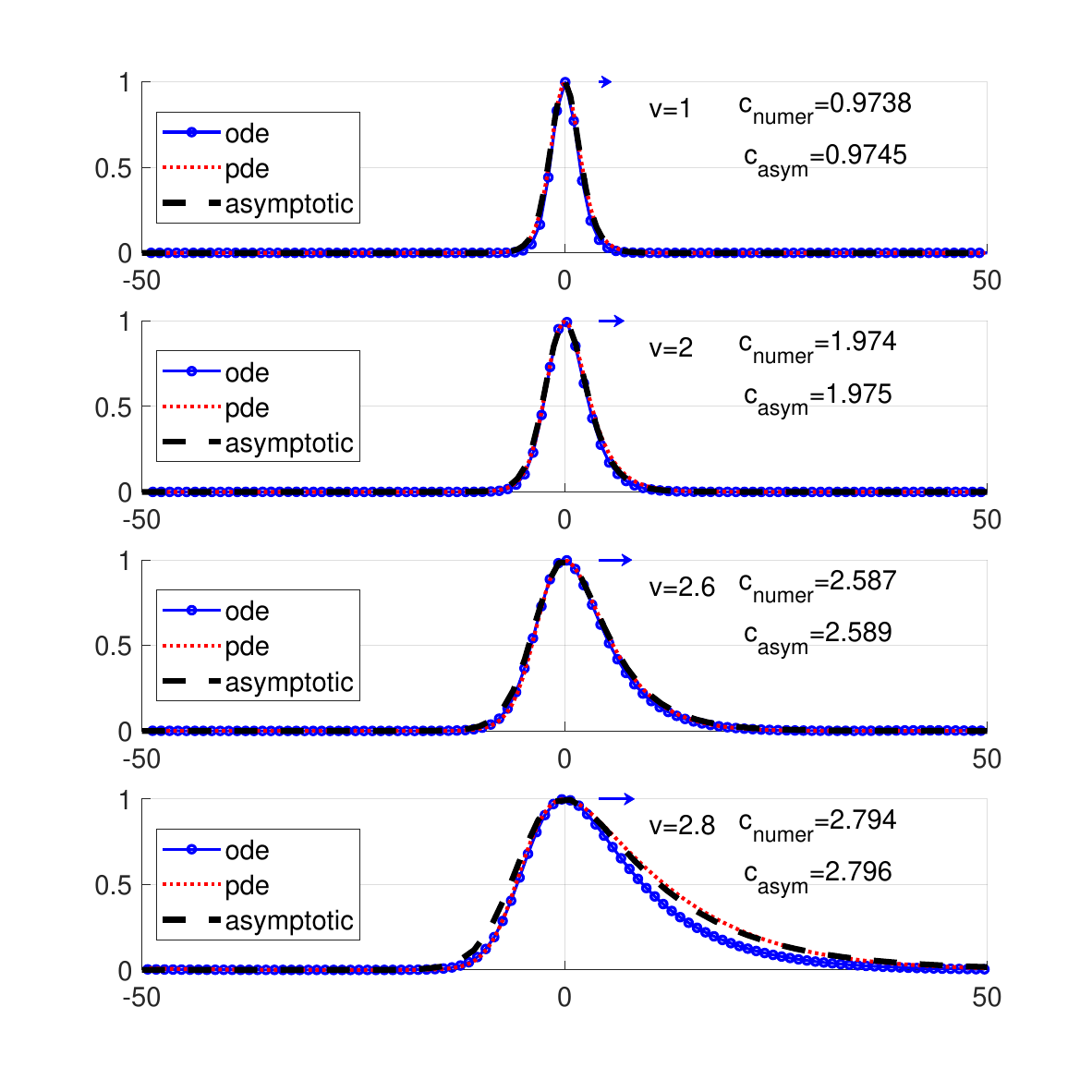}
\end{center}
\caption[Single cell model simulations]{Model I: 
Comparison between the discrete ODE
model (\ref{ode}), the PDE approximation (\ref{pde}) and the asymptotic
solution (\ref{uas}). For the ODE model the parameters were taken to
be:\ $a=1,\ b=5,\ N=100$ bins on the domain of size $L=100$ so that $h=1$, and
$v$ as indicated;  For the PDE model, we used implicit-explicit Euler scheme 
(implicit on diffusion and biharmonic terms, explicit on all other terms)
with $n=150$ meshpoints on the domain of size $[0,100]$. All profiles are
recentered on the pulse maximum, and the height is normalized to unity
(since the equations are invariant under scaling of $u$, scaling it
does not change the speed or profile). Periodic boundary conditions were used
for both ODE and PDE models.For each frame, $c_{\text{numer}}$ is the
speed of the pulse propagation, computed from the direct numerical simulations
of the ODE\ by tracking pulse location, whereas $c_{asym}$ is the asymptotic
prediction as given by the formula (\ref{c}).}%
\label{fig:model1}%
\end{figure}Equation (\ref{pde}) generalizes the model derived in
\cite{chavy2016local}, as it contains additional terms $-\beta u_{x}-\gamma
u_{xxx}$ which induce a pulse motion. We now generalize the method outlined in
\cite{chavy2016local} to derive an exact solution of (\ref{pde}) moving at a
constant speed $\tilde{c}$ to be determined. We start with a travelling pulse
ansatz of the form $u=u(x-\tilde{c}t).$ Then (\ref{pde}) reduces to an ODE%
\[
-\tilde{c}u_{y}=-u_{yy}-u_{yyyy}+\alpha\left(  \frac{u_{y}u_{yy}}{u}\right)
_{y}-\beta u_{y}-\gamma u_{yyy};\ \ \ \ y=x-\tilde{c}t.
\]
Integrate once and assume $u$ decays at $\infty.$ We obtain
\[
u^{\prime\prime\prime}=\left(  -\beta+\tilde{c}\right)  u-u^{\prime
}-\gamma u^{\prime\prime}+\alpha\frac{u^{\prime\prime}u^{\prime}}{u}.
\]
Because of the scaling symmetry $u\rightarrow\lambda u,$ we get a reduction of
order using the transformation%
\[
u=\exp(v);\ \ v^{\prime}=z
\]
to obtain%
\[
z^{\prime\prime}=-z+(\alpha-3)zz^{\prime}+(\alpha-1)z^{3}-\beta+\tilde
{c}-\gamma\left(  z^{2}+z^{\prime}\right)  ,\ \ \ z=u_{y}/u.
\]
We rewrite this as a system of ODEs,%
\begin{align}
\frac{dz}{dy}  &  =w;\\
\frac{dw}{dy}  &  =-z+(\alpha-3)zw+(\alpha-1)z^{3}-\beta+\tilde{c}%
-\gamma\left(  z^{2}+w\right)  .
\end{align}
We obtain a 1st order Abel\ ODE\ of the form%
\begin{equation}
\frac{dw}{dz}=\frac{-z}{w}+(\alpha-3)z+(\alpha-1)\frac{z^{3}}{w}+\frac
{-\beta+\tilde{c}}{w}-\gamma\left(  \frac{z^{2}}{w}+1\right)  .
\end{equation}
\begin{figure}[tb]
\begin{center}
\includegraphics[width=0.8\textwidth]{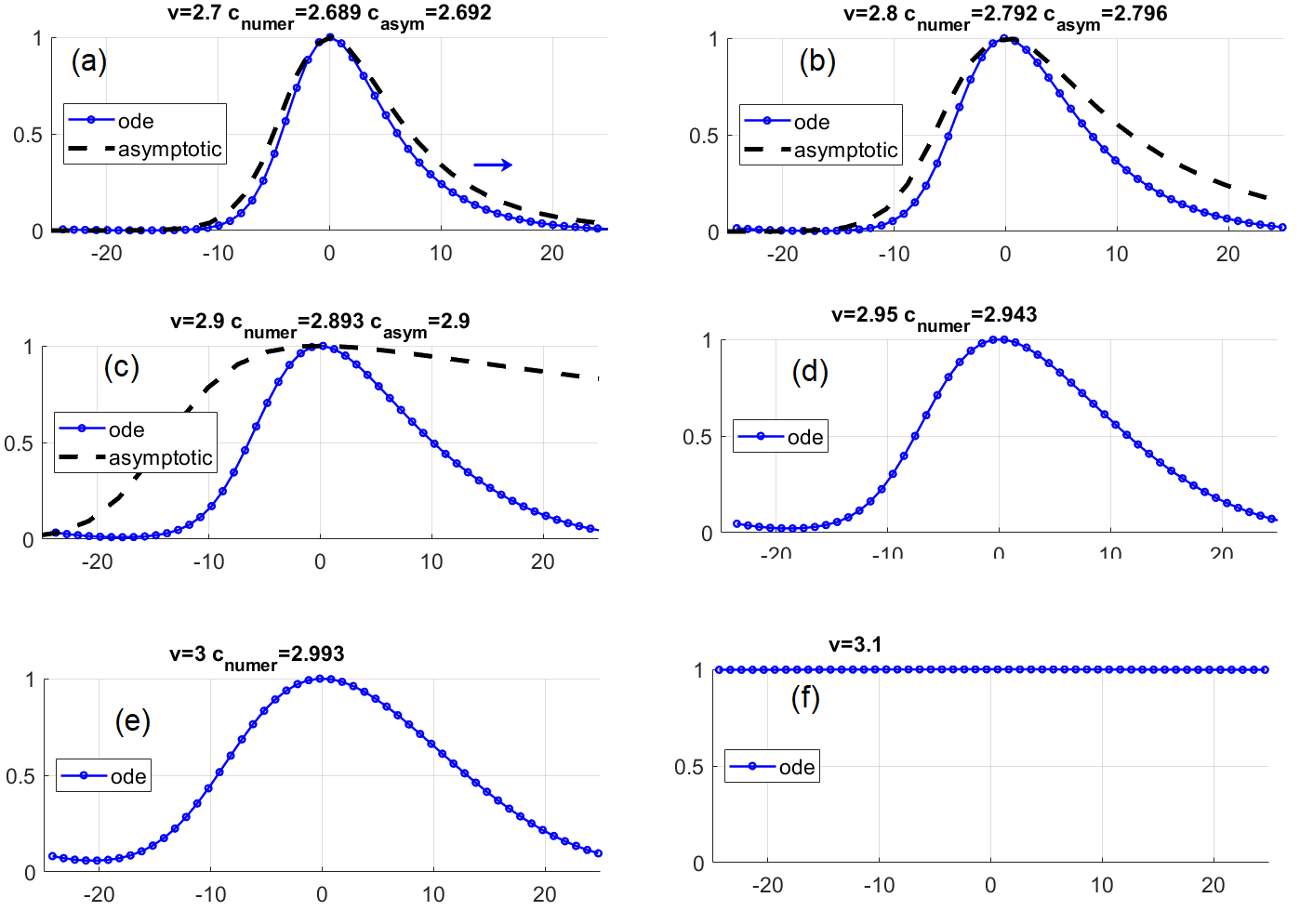}
\end{center}
\caption[Single cell model simulations]{Model I: 
Transition from a localized cluster
(when $v<v_{\ast}$) to a more broad non-homogeneous state when $v_{\ast
}<v<v_{c}$ and finally to a homogeneous constant state for $v>v_{c}.$ Here,
$a=h=d=1,\ b=5$ and $v$ is as indicated. 
All profiles have been rescaled to have a height of unity, and shifted to
have the peak at the origin. Blue line (``ode'') is the numerical
solution
to (\ref{ode}), at time $t=10000$. Black dashed line is the
asymptotic solution (2.17), where relevant.
For these parameter values, $v_{\ast
}=2.9099$ (given by\ (\ref{74}))\ and $v_{c}=3$. In (a)-(c), $v<v_{\ast}$ and
the localized cluster of the form (\ref{uas})\ exists, although the
approximation breaks down as $v\rightarrow v_{\ast}^{-}.$ In (d)-(e), the
localized state breaks down completely, and the breadth of the cluster is
comparible to the domain size. In (f), the solution eventually converges to a
homogeneous steady state.}%
\label{fig:vstar}%
\end{figure}

Next we use the ansatz\
\begin{equation}
w=A_{2}z^{2}+A_{1}z+A_{0}.
\end{equation}
This ansatz is self consistent if and only if%
\begin{align}
A_{2}  &  =\frac{\alpha-1}{2},\ \ \ A_{1}=-\frac{\left(  \alpha+1\right)
\gamma}{\alpha+3},\ \ \ A_{0}=-\frac{1}{2}+\frac{\left(  \alpha+1\right)
\gamma^{2}}{\left(  \alpha+3\right)  ^{2}},\nonumber\\
\tilde{c}  &  =\beta-\frac{\gamma}{\alpha+3}+\frac{2\left(  \alpha+1\right)
\gamma^{3}}{\left(  \alpha+3\right)  ^{3}}. \label{ctilde}%
\end{align}

Equation (\ref{ctilde}) gives the travelling speed. To obtain the pulse
profile, we unwind the transformations. Write%
\begin{equation}
\frac{dz}{dy}=A_{2}z^{2}+A_{1}z+A_{0}=c_{1}\left(  \left(  z+c_{2}\right)
^{2}-c_{3}^{2}\right)  , \label{2237}%
\end{equation}
where%
\begin{equation}
c_{1}=A_{2},\ \ c_{2}=\frac{A_{1}}{2A_{2}},\ \ c_{3}=\sqrt{\frac{A_{1}^{2}%
}{4A_{2}^{2}}-\frac{A_{0}}{A_{2}}}. \label{1559}%
\end{equation}
Integrating (\ref{2237}) then yields%
\[
z=-c_{2}-c_{3}\tanh\left(  c_{1}c_{3}y\right),
\]
so that $u=\exp(\int z),$ and hence
\begin{equation}
u=C\exp\left(  -c_{2}y\right)  \left(  \operatorname{sech}\left(  yc_{1}%
c_{3}\right)  \right)  ^{1/c_{1}}. \label{uas}%
\end{equation}
The constant $C$ is fixed by the total mass, as set by initial conditions.

When $v=0$, we find $c_{2}=0$, $\tilde{c}=0$, and (\ref{uas}) reduces to the
stationary symmetric sech profile as derived in \cite{chavy2016local}.

\bigskip

Going back to the the original variables $(x,t)$ in (\ref{xchange}), the
original speed $c$ is related to $\tilde{c}$ via%
\begin{equation}
c=\tilde{c}A^{3/2}B^{-1/2}%
\end{equation}
and the moving cluster profile is given by (\ref{uas}) with $y=\sqrt
{A/B}(x-ct).$ In the special case $a=1,d=1,h=1$ we have%
\begin{equation}
c=\frac{-43v^{4}+(1735b-194)v^{3}-27(73b-6)(11b-2)v^{2}+27(11b-2)^{2}%
(31b-2)v}{81(11b-2-v)^{3}} \label{c}%
\end{equation}

To validate formula (\ref{c}), take $a=1,b=5,d=1,h=1.$ In this case, the
cluster forms (i.e. the uniform equilibrium is unstable) provided that $0\leq
v<3$ (refer to (\ref{vc})). Figure \ref{fig:model1} shows a comparison between
the asymptotics, the original ODE model (\ref{ode}) as well as the PDE
approximation\ (\ref{pde}) for several values of $v$ in that range. Note that
the cluster becomes more asymmetric and wider as $v\rightarrow3^{-}.$ For
$v\geq3$, the solution converges to a uniform density, as predicted, even
starting with a cluster-type initial condition. Finally, we observe the
excellent predictive power of the formula (\ref{c}) for cluster speed $c$
(denoted by $c_{\text{asym}}$ 
in the figures); it accurately recovers the actual
observed speed in the ODE simulations (denoted by $c_{\text{numer}}$ in the
figures) across the entire range of $v$.

It is rather intriguing that the speed $c$ given by (\ref{c}) is very close
not only to the numerically computed value, but also to $v$ itself. In other
words, the cluster as a whole moves with a speed that is very close to the
individual particle speed. In fact, the biggest relative difference between
$c$ and $v$ occurs for small values of $v$, in which case (\ref{c}) reduces to%
\begin{equation}
c=\frac{31b-2}{33b-6}v+O(v^{2}). \label{1504}%
\end{equation}
With $b=5$ as above, the biggest deviation is 153/159, or 3.8\%.

Note that for $y\rightarrow+\infty$, we have $u\sim C\exp\left(  -\left(
c_{2}+c_{3}\right)  y\right)  $, whereas for $y\rightarrow-\infty,\ u\sim
C\exp\left(  -\left(  -c_{2}+c_{3}\right)  \left(  -y\right)  \right)  .$ It
follows that $u$ decays at $\pm\infty$ if and only if $c_{3}>\left\vert
c_{2}\right\vert $. This happens if and only if $A_{0}/A_{2}<0.$ Since
$A_{2}>0$, this is equivalent to $A_{0}<0$ or%
\begin{equation}
\left(  \alpha+3\right)  ^{2}>2\left(  \alpha+1\right)  \gamma^{2}. \label{73}%
\end{equation}

When $a,d,h=1,$ condition (\ref{73}) is equivalent to $v<v_{\ast}$, where
$v_{\ast}$ satisfies%

\begin{equation}
19v_{\ast}^{3}+\left(  -469b+146\right)  v_{\ast}^{2}+27\left(  11b-2\right)
\left(  13b-6\right)  v_{\ast}-27\left(  b-2\right)  \left(  11b-2\right)
^{2}=0. \label{74}%
\end{equation}
For example when $b=5,$ we obtain $v_{\ast}=2.9099$, whereas the threshold for
linear stability of a constant state is $v_{c}=3>v_{\ast}.$ Consequently, a
localized travelling pulse (which decays to zero as $y\rightarrow$ $\pm
\infty)$ exists only for $0<v<v_{\ast}<v_{c}.$ When $v>v_{c},$ the constant
state is stable. In the narrow regime $v_{\ast}<v<v_{c},$ the constant state
is unstable but the fully localized solution of the form (\ref{uas}) also does
not exist. Instead, in that narrow band we observe non-constant travelling
states which do not decay exponentially to zero. This is illustrated in Figure
\ref{fig:vstar}. More generally, Figure \ref{fig:phase-diag}(a)\ shows the
overall bifurcation diagram of $b$ vs. $v$. It appears that the narrow band
$v_{\ast}<v_{c}$ (labelled \textquotedblleft gap\textquotedblright\ in the
figure) exists for all $v_{c}>0$ (and indeed, from (\ref{74}), $v_{\ast}%
=v_{c}=0$ when $b=2$).

\section{Model II:\ biased aggregation.}

Instead of introducing an extra term for directed motion in (\ref{ode}), an
alternative is to incorporate direction bias directly into the aggregation
rate. Suppose that rather than choosing a neighbour at random, the cell
prefers to choose the neighbours to the right (for example due to light
gradient \textquotedblleft illuminating\textquotedblright\ the right side more
than the left side,
or because the cells are initially polarized towards the right). 
To simplify the computations, we will assume the sensing
radius $d$ to be $d=1.$ We then modify the model (\ref{ode}) as follows:%
\begin{align}
\frac{dU_{j}}{dt}  &  =a\left(  U_{j-1}-2U_{j}+U_{j+1}\right)  +b\left(
\eta_{j+1}^{-}U_{j+1}-U_{j}+\eta_{j-1}^{+}U_{j-1}\right)  ;\label{ode2}\\
\eta_{j}^{+}  &  =\frac{\left(  1+hr\right)  U_{j+1}}{\left(  1+hr\right)
U_{j+1}+\left(  1-hr\right)  U_{j-1}},\ \ \eta_{j}^{-}=\frac{\left(
1-hr\right)  U_{j-1}}{\left(  1+hr\right)  U_{j+1}+\left(  1-hr\right)
U_{j-1}}.
\end{align}

The new parameter $r$ represents the bias towards choosing a neighbour to the
right, and $h=L/n$ is the bin width (assumed to be small). Note that $\eta
_{j}^{+}+\eta_{j}^{-}=1$ still holds. As before, we expand to fourth order in
grid spacing $h$, obtaining the PDE%

\begin{equation}
u_{t}=\left[  au_{xx}+b\left(  -\frac{u_{xx}}{2}-ru_{x}\right)  \right]
h^{2}+\left[  a\frac{u_{xxxx}}{12}+b\left\{  \left(  -\frac{7u_{xxx}}%
{24}+\frac{u_{xx}u_{x}}{2u}\right)  _{x}+\left(  -\frac{u_{xx}}{6}+\frac
{u_{x}^{2}}{u}\right)  _{x}r+u_{xx}r^{2}\right\}  \right]  h^{4}%
\end{equation}

\begin{figure}[ptbh]
\begin{center}
\includegraphics[width=0.8\textwidth]{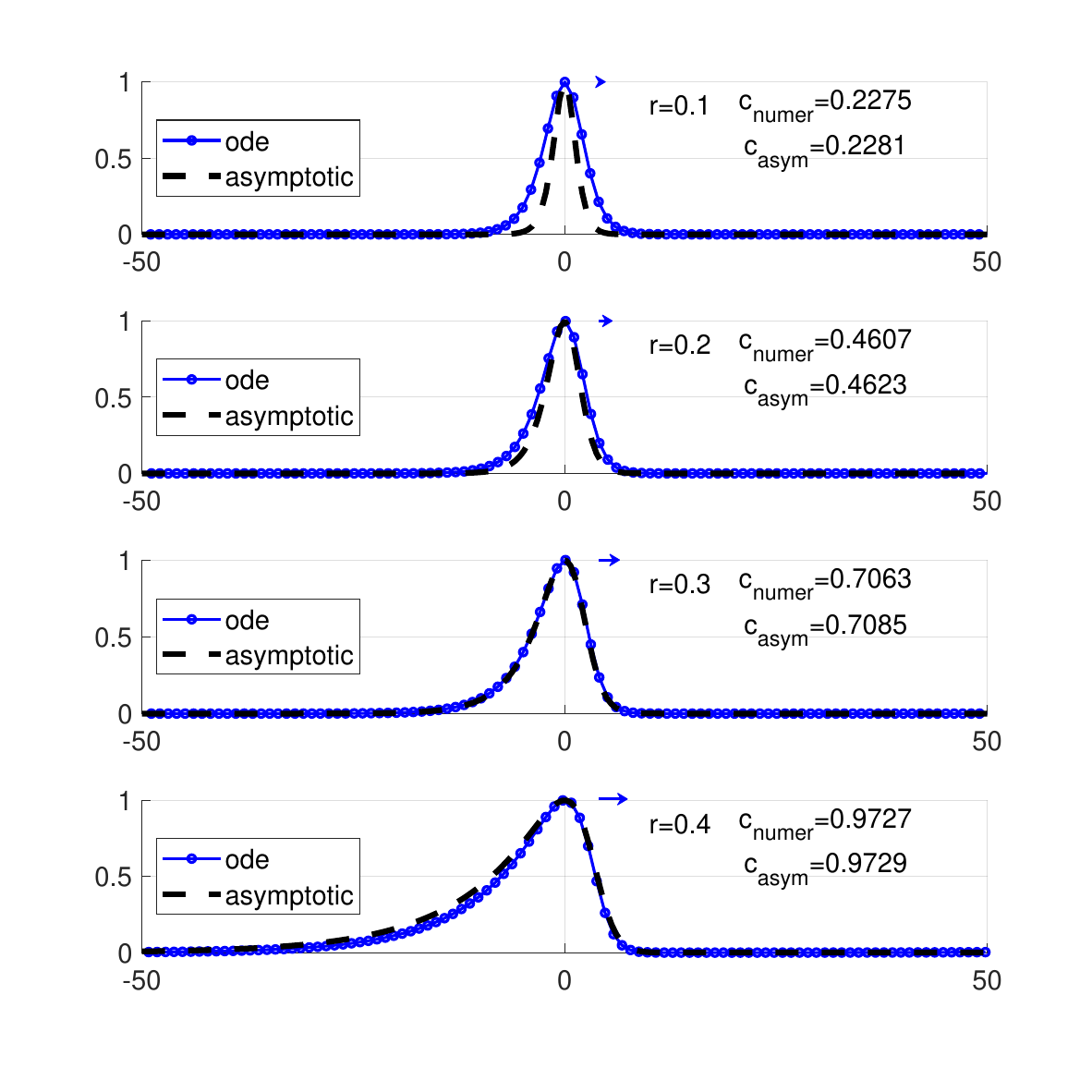}
\end{center}
\caption[Single cell model simulations]{Model II: 
Comparison between the discrete ODE
model (\ref{ode2}) and the asymptotic solution
(\ref{A0model2}-\ref{1505}, \ref{1559}, \ref{uas}). For
the ODE model, the parameters were taken to be:\ $a=1,\ b=2.5,\ N=100,\ L=100$,
$h=1$, with periodic boundary conditions and $r$ as indicated. For each frame,
$c_{\text{numer}}$ is the speed of the pulse propagation, computed from the
direct numerical simulations by tracking pulse location, whereas $c_\text{asym}$
is the asymptotic prediction as given by the formula (\ref{1505}). The last
frame $(r=0.4)$ corresponds to a bistable region (see also Figure 6).}%
\label{fig:model2}%
\end{figure}Although more complex than the PDE (\ref{pde}), the analysis
follows along similar lines. In particular, linear analysis shows that the
homogeneous state is unstable provided that%
\begin{equation}
a+b\left(  -\frac{1}{2}+h^{2}r^{2}\right)  <0. \label{y6541}%
\end{equation}
By analogy with Model I, we define the critical value $r_{c}$ to be the root
of (\ref{y6541}):%
\begin{equation}
r_{c}=\frac{1}{h}\left(  1/2-a/b\right)  ^{1/2}. \label{rc}%
\end{equation}
The homogeneous state is unstable when $\left\vert r\right\vert <r_{c},$ and
numerics indicate that clusters form when this instability is triggered.

Following \S \ref{sec:1}, we seek a travelling cluster solution of the form
\[
u(x,t)=u(y),\ \ y=x-ct,
\]
corresponding to a cluster travelling to the right with a constant speed $c.$
The procedure is then the same as in \S \ref{sec:1}. Upon integrating once, we
obtain a 3rd order ODE%
\begin{equation}
0=\left(  au_{y}+\left(  -ru-\frac{u_{y}}{2}\right)  b\right)  h^{2}+\left[
a\frac{u_{yyy}}{12}+b\left\{  \left(  -\frac{7u_{yyy}}{24}+\frac{u_{yy}u_{y}%
}{2u}\right)  +\left(  -\frac{u_{yy}}{6}+\frac{u_{y}^{2}}{u}\right)
r+u_{y}r^{2}\right\}  \right]  h^{4}+cu
\end{equation}

Substitute $u=e^{v};$ \ \ $v^{\prime}=z,$ $\frac{dz}{dy}=w$, $\frac{dw}%
{dy}=\frac{dw}{dz}w$, to obtain a first order ODE%
\begin{align*}
0  &  =h^{4}w\left(  2a-7b\right)  \frac{dw}{dz}+h^{4}\left(  2a+5b\right)
z^{3}+20b\,h^{4}r\,z^{2}+3h^{4}\left(  2a-3b\right)  wz\\
&  +12h^{2}\left(  2b\,h^{2}r^{2}+2a-b\right)  z-4b\,h^{4}rw-24b\,h^{2}r+24c.
\end{align*}

We now substitute $w=A_{2}z^{2}+A_{1}z+A_{0}$, to obtain a polynomial system
relating $A_{0},A_{1},A_{2}$ and $c.$ It is fully solvable, having two
solutions. One, corresponding to $A_{2}=-1$, is non-physical, while the other has%

\begin{align}
A_{0}  &  =\frac{4b\left(  108a^{2}-572ab+1447b^{2}\right)  r^{2}}{9\left(
7b-2a\right)  \left(  2a-11b\right)  ^{2}}+\frac{6\left(  2a-b\right)
}{\left(  7b-2a\right)  h^{2}},\ \label{A0model2}\\
A_{1}  &  =\frac{4b\left(  22a-65b\right)  r}{3\left(  7b-2a\right)  \left(
2a-11b\right)  },\ \ A_{2}=\frac{2a+5b}{2\left(  7b-2a\right)  }%
,\label{Amodel2}\\
c  &  =-\frac{28b^{2}h^{4}\left(  108a^{2}-572ab+1447b^{2}\right)  r^{3}%
}{81\left(  2a-11b\right)  ^{3}}-\frac{h^{2}b\left(  22a+19b\right)
r}{3\left(  2a-11b\right)  }. \label{1505}%
\end{align}

This determines the speed $c$, as well as the corresponding cluster profile
given by (\ref{1559}, \ref{uas}). Figure \ref{fig:model2} shows a direct
 comparison between numerical simulations of the ODE\ model
(\ref{ode2})\ and the asymptotic profile as well as group velocity
(\ref{1505}). Excellent agreement is observed.

\begin{figure}[ptbh]
\begin{center}
\includegraphics[width=0.98\textwidth]{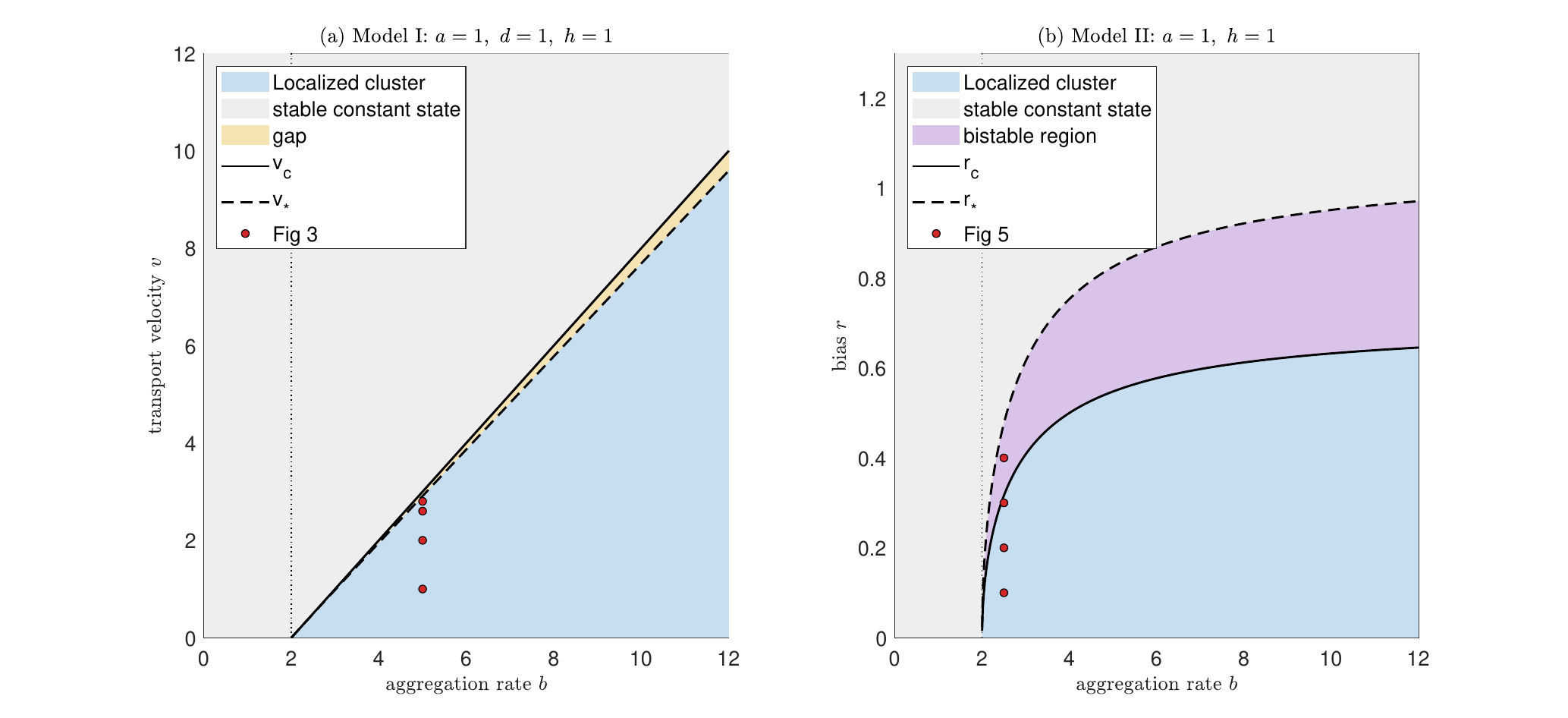}
\end{center}
\caption{ (a) Bifurcation diagram for Model I, in the $(b,v)$ plane, with
$a=h=d=1$. Constant steady state is stabe in the grey region $v>v_{c}$.
Localized solutions exist in the blue region $v<v_{\ast}$. Since $v_{\ast
}<v_{c},$ there is a \textquotedblleft gap\textquotedblright\ region where the
constant state is unstable but fully localized clusters do not exist (see also
Figure 4). Red dots denote the simulation parameters from Figure 3.
(b)\ Bifurcation diagram for Model II, in the $(b,r)$ plane with $a=h=d=1$.
Constant steady state is stable in the grey region $r>r_{c}$. Localized
solutions exist in the blue region $r<r_{\ast}$. Since $r_{\ast}>r_{c},$ there
is a bistable\ region where constant stable state coexists with a localized
cluster state. Red dots denote the simulution parameters from Figure 5.}%
\label{fig:phase-diag}%
\end{figure}

As in Model I, a localized cluster ceases to exist when $A_{0}=0$, corresponding
to the value of $r=r_{\ast}$ given by%
\begin{equation}
r_{\ast}=\frac{1}{h}\sqrt{\frac{27\left(  b-2a\right)  \left(  2a-11b\right)
^{2}}{2b\left(  108a^{2}-572ab+1447b^{2}\right)  }}. \label{rs}%
\end{equation}
(The cluster exists when $\left\vert r\right\vert <r_{\ast}$ and does not
exist when $\left\vert r\right\vert >r_{\ast}$). In the example from Figure 5
where $a=1,b=2.5,h=1,$ we obtain $r_{\ast}=0.4768$ and $r_{c}=0.3162,$ so that
$r_{\ast}>r_{c}.$ We have the following situation:\ when $r<r_{c},$ the
homogeneous state is unstable and localized cluster exists. When $r>r_{\ast},$
the homogeneous state is stable and cluster does not exist. When
$r_{c}<r<r_{\ast},$ the cluster exists and the homogeneous state is stable.
Numerics further indicate that the cluster is indeed stable in this band once
formed, see the frame for $r=0.4$ in Figure 5. As such, $r\in\left(
r_{c},r_{\ast}\right)  $ is a region of bistability where both cluster and
homogeneous state can be observed depending on the initial conditions. Figure
\ref{fig:phase-diag}(a)\ shows the overall bifurcation diagram of $b$ vs. $r$;
the bistability band $r_{c}<r_{\ast}$ exists for all $b>2a.$

\section{Discussion}

We have generalized as well as simplified the model proposed in
\cite{msel1,chavy2016local} to incorporate directed motion of cells. We
proposed two models that induce group motion, either by adding simple directed
motion of cells (Model I), or by adding a biased sensing mechanism (Model II).
In the limit of a large number of particles, we derived an analytical
expression for both the aggregation profile as well as its group velocity.
This generalizes results in \cite{chavy2016local}, where the model leads to a
stationary spike (zero velocity). The advantage of these models is the
relative simplicity of the aggregation mechanism, and availability of the full
analytical description of the cluster profile as well as its velocity in the
mean-field limit. This is one of the very few models where such exact and full
description is possible.

Let us contrast the two models. In Model I, each cell is assigned an intrinsic
velocity $v.$ We found that although the overall group velocity of the cluster
$c$ was definitely not equal to individual cell velocity $v$, it was
suprisingly close (within 4\% or less for all the cases we studied, see e.g.
(\ref{1504})). By contrast, Model II\ does not dictate intrinsic particle
velocity, and instead assigns a bias $r$, so that it \textquotedblleft
weights\textquotedblright\ its neighbours to the right more than its
neighbours to the left. The overall group velocity is a nonlinear
(cubic)\ function of the bias $r$ that we explicitly computed in (\ref{1505}).

Another distinction is the profile of the right-moving cluster shown in
Figures \ref{fig:model1}, \ref{fig:model2}. The cluster in Model I is steeper
on the left and shallower on the right. The opposite is true for Model II.
Such asymmetry is present in many models of cluster motion, but usually the
cluster is sharper on the right (in the direction of motion, as in Model II).
This is the case, for instance, in the Klausmeier model
\cite{sherratt2011pattern}, as well as in many cellular aggregation models
\cite{smsp}. This suggests that Model II\ may be more relevant for travelling
biological clusters.

An interesting new phenomenon is the existence of a \textquotedblleft
gap\textquotedblright\ window in Model I (where neither uniform state is
stable, nor the localized cluster exists), and bistablity window in Model II
(where both the uniform state is stable and localized cluster exists)\ -- see
Figure 6. Both of these phenomena are specific to moving pulses: no such
windows exist in the original model \cite{chavy2016local}. Within the narrow
\textquotedblleft gap\textquotedblright\ window, numerics show a novel type of
solutions which are neither localized, nor uniform, but span the entire domain
(Figure 4). Further insights into this region is an interesting open question.

Our results have several physical implications. First, in Model I, we found 
that decreasing the aggregation strength, or increasing the fluid flow past some 
critical threshold $v_c$ flattens out the cluster, and eventually disables 
aggregate cohesion. This is consistent with experimental observations that 
microfluidic flow strength can toggle between modes of cell behaviour (such as amoeboid vs mesenchymal \cite{huang2015interstitial}) or, in our model, single vs collective motion. 
As far as cell types, larger random motility or lower sensing radius in the cells
 lowers the critical flow velocity for transition, making aggregates less stable. 
 Hence, to form stable aggregates, the cells should have stronger 
 cell-cell attraction, sense neighbours over a larger distance, and reduce 
 their random exploration. In contrast, in Model II, we found that increasing 
 the cells' biased motion (e.g., faster polarized cells) makes the cluster less 
 symmetric, with a longer shallower ``tail". Eventually, straggler cells are 
 lost from the back so that the cluster cannot be maintained. In other words,
 if cells tend to move too rapidly in a biased direction, neighbours behind them 
 cannot keep up. This model helps to understand the behaviour of ``cell trains'' 
 moving along 1D adhesive strips, as in \cite{jain2020role},
  where a transition from
 single cells to a band of cells can be observed.

\bigskip

\section{Acknowledgements}

Claude AI was shown an early draft of this paper. It discovered
and sketched the \textquotedblleft gap\textquotedblright\ and
\textquotedblleft bistable\textquotedblright\ regions in Figure 6. The
authors independently verified the correctness of the figure.

\bibliographystyle{elsarticle-num}
\bibliography{bib}

\end{document}